\documentclass[aps,prb,twocolumn]{revtex4}
\pdfoutput=1
\usepackage[latin9]{inputenc}
\usepackage{graphicx}
\usepackage{amsmath}
\usepackage{amssymb}

\begin{document}

\title{Recent progress in open quantum systems: Non-Gaussian noise and decoherence in fermionic systems}


\author{
  Clemens Neuenhahn,
  B. Kubala,
  B. Abel, 
  Florian Marquardt}

\affiliation{Department of Physics, Center for NanoScience, and Arnold Sommerfeld
Center for Theoretical Physics, Ludwig Maximilians Universit\"at M\"unchen,
Theresienstr. 37, 80333 Munich, Germany}

 
\begin{abstract}

We review our recent contributions to two topics that have become
of interest in the field of open, dissipative quantum systems: non-Gaussian
noise and decoherence in fermionic systems. Decoherence by non-Gaussian
noise, i.e. by an environment that cannot be approximated as a bath
of harmonic oscillators, is important in nanostructures (e.g. qubits) where there
might be strong coupling to a small number of fluctuators. We first revisit the pedagogical example of 
dephasing by classical telegraph noise. Then we address two models where the quantum nature of the noise
becomes essential: "quantum telegraph noise" and dephasing by electronic shot noise. In fermionic
systems, many-body aspects and the Pauli principle have to be taken
care of when describing the loss of phase coherence. This is relevant  in electronic quantum transport through metallic and semiconducting structures. Specifically, we recount our recent results regarding dephasing in a chiral interacting electron liquid, as it is realized in the electronic Mach-Zehnder interferometer. This model can be solved employing the technique of bosonization as well as a physically transparent semiclassical method.
{\em - Manuscript submitted to the proceedings of the XXXII International Conference on Theoretical Physics, "Coherence and Correlations in Nanosystems", Ustron, Poland, September 2008 [subm. to physica status solidi (b)]}
\end{abstract}

\maketitle

\section{Introduction}

The coupling of a quantum system to a noisy environment leads to decoherence \cite{2000_Weiss_QuantumDissipativeSystems,1990_04_SAI,1991_Zurek_PhysicsToday,2002_Breuer_Book},
i.e. the loss of quantum-mechanical phase coherence and the suppression
of the associated interference effects. Understanding decoherence
is interesting for fundamental reasons (the quantum-classical crossover,
the measurement problem, etc.), and it is essential for achieving
the long dephasing times necessary for building a quantum computer
and other applications. 

In this brief review we present two topics that have become of recent
interest in the field of quantum dissipative systems: The first is
decoherence by 'non-Gaussian noise', that is, environments that cannot
be described by the usual bath of harmonic oscillators and give rise
to qualitatively new features. This is relevant especially for qubit decoherence
in nanostructures. The second topic concerns decoherence in
many-fermion systems, where issues such as Pauli blocking have to
be taken into account. These are important to discuss decoherence
in solid-state electronic transport interference experiments.

\section{Dephasing by non-Gaussian noise}

In the following, we will first present in some detail the pedagogical
example of dephasing by classical telegraph noise, where the most
important features of dephasing by non-Gaussian noise can be discussed
in an exact solution.
Then we review two recent works dealing with quantum non-Gaussian
noise, both in equilibrium (dephasing of a qubit by a fluctuator),
and out of equilibrium (dephasing by electronic shot noise).

\begin{figure}
\includegraphics[width=1\columnwidth]{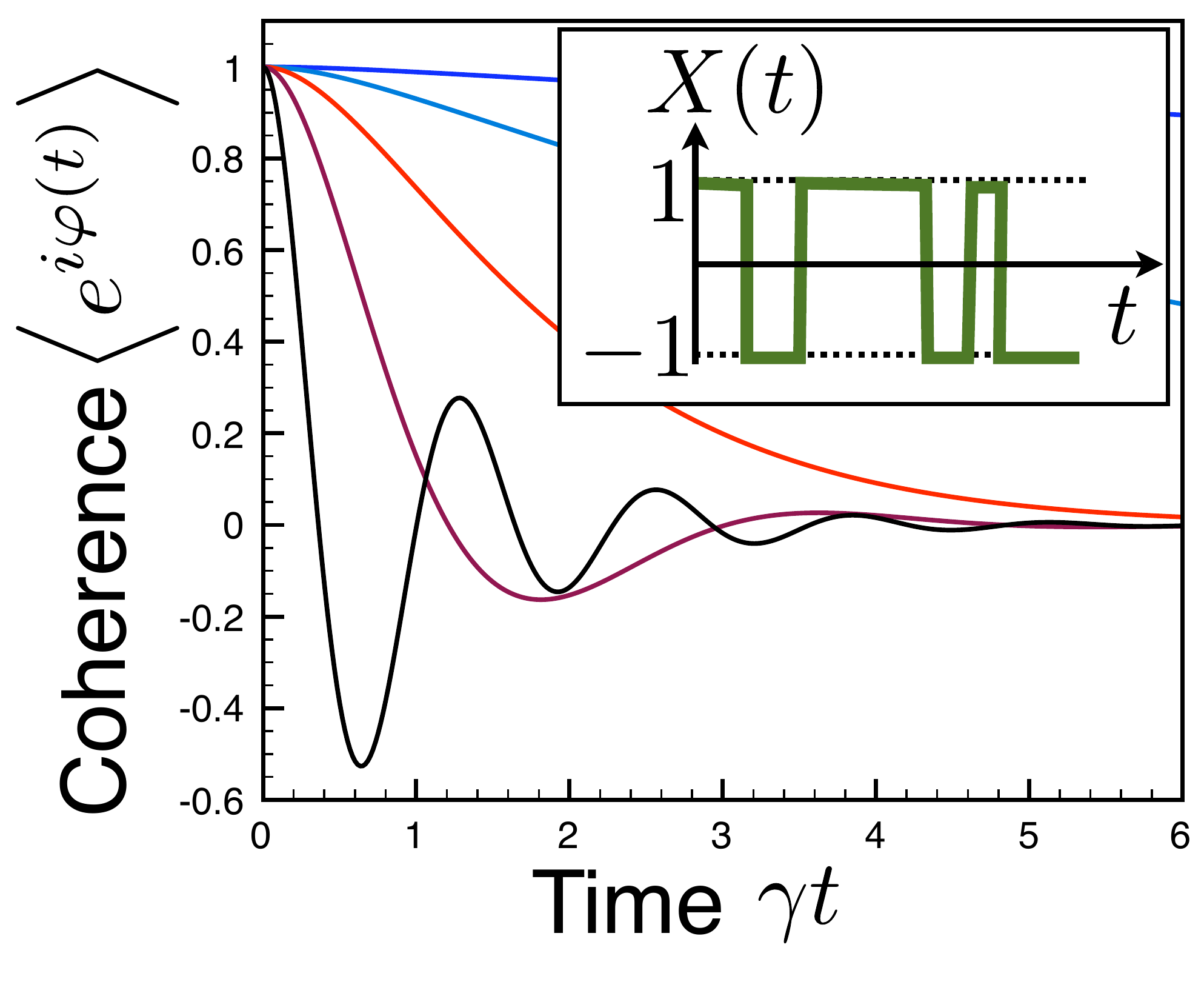}

\caption{\label{classfig}The coherence of a qubit subject to pure dephasing
by classical telegraph noise (inset), for increasing coupling strength
(top to bottom): $\alpha/\gamma=0.2,0.5,1,2.0,5.0$. Note the oscillations
beyond the strong-coupling threshold $\alpha/\gamma=1$.}

\end{figure}

\subsection{Classical telegraph noise as an example of non-Gaussian noise}

Most discussions of dephasing are concerned with harmonic oscillator
environments or their counterparts, Gaussian noise processes. Not
only are these the simplest class of environments, but they also do
describe many important physical examples. In addition, they often
turn out to be good approximations even in cases where the microscopic
Hamiltonian of the environment is not a sum of harmonic oscillators.
In the case of noise processes, this can be understood essentially
from the central limit theorem: If the process is a sum of many independent
fluctuating contributions, then that sum will be Gaussian to a good
approximation, even if the individual terms are not. 

However, there are important exceptions, and a part of current research
into quantum dissipative systems is directed towards such situations. 

Here, we will illustrate the salient features by reviewing the simplest
pedagogical example of a non-Gaussian noise process, namely telegraph
noise. This is the name for a process $V(t)$ that can take only two
values and randomly jumps between those values. Then, the distribution
of possible values $V(t)$ at any given time $t$ is obviously not
a Gaussian. The probability for these jumps to occur in a given time-interval
is assumed to be independent of the previous history of the process,
i.e. the process is of the {}``Markov'' type. 

There are many possible realizations for such a process, among them
an electron tunneling incoherently between two different locations
inside a crystal. Such an electron gives rise to an electric field
that can shift the energy of another coherent quantum two-level system.
The energy shift $V(t)$ of the two-level system will depend on the
current position of the electron, and is thus of the random telegraph
type. A qubit subject to these energy fluctuations will acquire a
random phase, $\varphi(t)=-\hbar^{-1}\int_{0}^{t}V(t')dt'$, such
that its coherence (the off-diagonal entry of the density matrix)
is suppressed by the factor $\left\langle e^{i\varphi}\right\rangle $.
It is our aim to describe the time-evolution of the qubit's coherence.

Let us, for simplicity, assume that the jump rates between the two
states are equal to $\gamma$, and the two possible values $V=\pm\hbar\alpha=\hbar\alpha X$
(with $X=\pm1$) thus occur with equal probability. (It is easy to
generalize the results below to the case of unequal rates). The correlation
function for such a process is

\begin{equation}
\left\langle V(t)V(0)\right\rangle =\hbar^{2}\alpha^{2}e^{-2\gamma|t|}\,.\label{PD-TelegraphCorrelator}\end{equation}
If $V(t)$ were a Gaussian process, this correlator alone would be
sufficient to calculate the decay of the coherence. A Gaussian process
with the correlator (\ref{PD-TelegraphCorrelator}) is of the Ornstein-Uhlenbeck
type. We would obtain:

\begin{equation}
\left\langle e^{i\varphi(t)}\right\rangle =\exp\left[-\frac{\alpha^{2}}{2\gamma}(t-\frac{1}{2\gamma}(1-e^{-2\gamma t}))\right]\,.\label{PD-GaussianTelegraph}\end{equation}
In that case, the decay of the coherence is monotonous in time. However,
we will see that the true telegraph noise process induces a much more
interesting behaviour. In the end, we will compare the coherences
predicted by the two models.

In order to find the time-evolution of $\left\langle e^{i\varphi}\right\rangle $,
we can express it through the time-evolution of the probability densities
$p_{+}(\varphi,t)$ and $p_{-}(\varphi,t)$ that describe the probabilities
of finding the telegraph process in state $+$ or $-$ and having,
at the same time, a certain value $\varphi$ of the phase. The equations
governing the time-evolution are of {}``Markov'' type, i.e. they
only depend on the current values of $p_{\pm}(\varphi,t)$:

\[
\dot{p}_{\pm}(\varphi,t)=\pm\alpha\partial_{\varphi}p_{\pm}(\varphi,t)+\gamma(p_{\mp}(\varphi,t)-p_{\pm}(\varphi,t))\]
The phase evolves deterministically, $\dot{\varphi}=-\alpha X$, if
the telegraph process is in the state $X=\pm1$. As a consequence,
the probability density is shifted to the left ($p_{+}$) or to the
right ($p_{-}$), which is reflected by the drift term $\pm\alpha\partial_{\varphi}p_{\pm}$
on the rhs. The second part describes the possibility of jumping,
with a rate $\gamma$, between those two states. Initially, both states
$\pm$ have equal probability and the phase is zero: $p_{+}(\varphi,t=0)=p_{-}(\varphi,t=0)=\frac{1}{2}\delta(\varphi)$.

However, instead of solving these equations for the probability densities
(which is easily done after Fourier transformation) we can take a
short-cut. The trick is to write down equations of motion for $\left\langle e^{i\varphi}\right\rangle $
and $\left\langle Xe^{i\varphi}\right\rangle $, and to observe that
these equations close. First, we find

\begin{equation}
\frac{d}{dt}\left\langle e^{i\varphi(t)}\right\rangle =i\left\langle \dot{\varphi}(t)e^{i\varphi(t)}\right\rangle =-i\alpha\left\langle X(t)e^{i\varphi(t)}\right\rangle \,.\label{eq:ephieq}\end{equation}
In the next step, we have to take the time-derivative of $\left\langle Xe^{i\varphi}\right\rangle $.
Here we can employ the fact that $X$ is a dichotomous process, i.e.
$X^{2}\equiv1$ is a constant. Therefore 

\begin{equation}
\left\langle X\frac{d}{dt}e^{i\varphi}\right\rangle =-i\alpha\left\langle X^{2}e^{i\varphi}\right\rangle =-i\alpha\left\langle e^{i\varphi}\right\rangle \, .\end{equation}
We are still left, however, with the problem of evaluating $\left\langle e^{i\varphi}\frac{d}{dt}X\right\rangle $.
We note first that the random process $dX/dt$ consists of a series
of delta peaks, with prefactors $\pm2$, corresponding to transitions
from $X=-1$ to $X=+1$ and vice versa. We can express this by writing
$dX/dt=-2XY$, where $ $$Y$ is a Poisson process, consisting of
uncorrelated delta peaks of weight $1$. The probability of observing
a jump (i.e. one of these peaks) inside a time interval around $t$
is independent of the previous history $Y(t')$ at times $t'<t$,
and thus also independent of $X$ and $\varphi$. Therefore 

\begin{eqnarray}
\left\langle e^{i\varphi}dX/dt\right\rangle =-2\left\langle e^{i\varphi}XY\right\rangle =\nonumber \\
-2\left\langle e^{i\varphi}X\right\rangle \left\langle Y\right\rangle =-2\left\langle e^{i\varphi}X\right\rangle \gamma &  & \,,\end{eqnarray}
and we arrive at

\begin{equation}
\frac{d}{dt}\left\langle X(t)e^{i\varphi(t)}\right\rangle =-i\alpha\left\langle e^{i\varphi(t)}\right\rangle -2\gamma\left\langle X(t)e^{i\varphi(t)}\right\rangle \,.\label{eq:secondeq}\end{equation}
Taking the derivative of Eq.~(\ref{eq:ephieq}), and inserting Eq.~(\ref{eq:secondeq}),
we find that $\left\langle e^{i\varphi}\right\rangle $ obeys the
equation of motion of a damped harmonic oscillator:

\begin{equation}
\frac{d^{2}}{dt^{2}}\left\langle e^{i\varphi}\right\rangle =-\alpha^{2}\left\langle e^{i\varphi}\right\rangle -2\gamma\frac{d}{dt}\left\langle e^{i\varphi}\right\rangle \,.\end{equation}
The solution of this equation (with the proper initial conditions
$\left\langle e^{i\varphi}\right\rangle =1$ and $\left\langle Xe^{i\varphi}\right\rangle =0$)
is:

\begin{equation}
\left\langle e^{i\varphi}\right\rangle =\frac{1}{2}e^{-\gamma t}\left\{ \left[1+\frac{\gamma}{i\Omega}\right]e^{i\Omega t}+\left[1-\frac{\gamma}{i\Omega}\right]e^{-i\Omega t}\right\} \,,\label{PD-coherenceTelegraphNoise}\end{equation}
with $\Omega\equiv\sqrt{\alpha^{2}-\gamma^{2}}$. This is the coherence
of a two-level system subject to pure dephasing by random telegraph
noise of strength $\alpha$ and switching rate $\gamma$ (see Fig.~\ref{classfig}).

For strong damping (weak coupling), $\gamma>\alpha$, the fictitious
harmonic oscillator is overdamped, $\Omega$ is imaginary, and the
decay of $\left\langle e^{i\varphi}\right\rangle $ is monotonous,
qualitatively similar to the expectation for dephasing by Gaussian
noise. In contrast, for strong coupling, $\alpha>\gamma$, the coherence
will display damped oscillations (i.e. it may become negative). This
behaviour is qualitatively different from what would have been predicted
in the case of a Gaussian random process (see Eq.~(\ref{PD-GaussianTelegraph})
above). In fact, it cannot be mimicked by \emph{any} Gaussian process,
since there the coherence can never become negative! The frequency
scale $\Omega$ of the oscillations becomes equal to the coupling
strength $\alpha$ in the limit of a vanishing switching rate $\gamma\rightarrow0$.
This can be understood quite easily. In that limit, the positive and
negative sign in $\varphi=\pm\alpha t$ each occurs with probability
$1/2$, thus $\left\langle e^{i\varphi}\right\rangle =\frac{1}{2}[e^{i\alpha t}+e^{-i\alpha t}]=\cos(\alpha t)$. 

In the particular limit of a very high switching rate, $\gamma/\alpha\rightarrow\infty$,
we expect the Gaussian process to be a good approximation: In that
limit, the phase is a sum over many small independent contributions
(from small time-intervals of order $\gamma^{-1}$) and thus should
become a Gaussian variable according to the central limit theorem,
performing a close approximation to a random walk. Indeed, evaluating
the exact result (\ref{PD-coherenceTelegraphNoise}) for the coherence
in the limit $\alpha/\gamma\rightarrow0,\,\,\alpha^{2}t/\gamma={\rm const}$
(which implies $\gamma t\gg1$), we find, to leading order in $\alpha/\gamma$:

\begin{equation}
\left\langle e^{i\varphi}\right\rangle \approx\left[1+\frac{\alpha^{2}}{4\gamma^{2}}\right]e^{-\frac{\alpha^{2}}{2\gamma}t}\,.\end{equation}
This coincides with the result for the corresponding Gaussian process,
Eq. (\ref{PD-GaussianTelegraph}), evaluated in the same limit and
to the same approximation.

In this example, we have thus found the coherence of a two-level system
for a non-Gaussian process and learned that it can deviate qualitatively
from that of a Gaussian process. Only in the limit of weak coupling,
when the effect on the phase during one correlation time $1/\gamma$
of the fluctuations is very small, one can obtain the same results
from a Gaussian process with the same correlator.

\subsection{Quantum telegraph noise}

We now want to ask about situations in which there is a quantum analogue
of classical telegraph noise. This is in the spirit of the concept
of 'Quantum Brownian motion' \cite{1983_CaldeiraLeggett_QuantumBrownianMotion},
where one asks for a quantum model which will yield classical Brownian
motion and velocity-proportional friction in the high temperature
limit. 

Possibly the simplest model is a single level onto which particles
may hop. If we forbid multiple occupation, either by postulating interactions
(hard), or by dealing with fermions and appealing to the Pauli principle
(simple), then the occupation number of that level will fluctuate
between $0$ and $1$. We have recently looked at the second case
\cite{2008_05_AbelMarquardt_QuantumTelegraphNoise}, in a model where
a single defect level is tunnel-coupled to an infinite reservoir of
non-interacting fermions. The charge fluctuations of this level couple
to the energy splitting of a qubit, thus giving rise to dephasing.
This is a realistic model for describing the dephasing of charge qubits
by two-level fluctuators, and various aspects and limiting cases of
that model had been studied before \cite{2002_06_PaladinoFazio_TelegraphNoisePRL,2005_08_Lerner_QuantumTelegraphNoiseLongTime,2006_03_GalperinAltshuler_NonGaussianQubitDecoherence,2006_01_SchrieflShnirman_DecoherenceTLS}. 

However, in \cite{2008_05_AbelMarquardt_QuantumTelegraphNoise} we
provide an exact solution and evaluate it numerically to discuss the
qubit dephasing at arbitrary times (not only in the long-time limit
that had been tackled in \cite{2005_08_Lerner_QuantumTelegraphNoiseLongTime}).
The off-diagonal element of the density matrix of the qubit is suppressed
by a factor $D(t)$ that may be written as a determinant in the single-particle
Hilbert space of the fermionic bath (including the defect level):

\begin{equation}
D(t)=\det\left(1-\hat{n}+e^{i(\hat{H}_{B}-\frac{v}{2}\hat{Q})t}e^{-i(\hat{H}_{B}+\frac{v}{2}\hat{Q})t}\hat{n}\right).\label{eq:determinant}\end{equation}
Here $\hat{n}=f(\hat{H}_{B})$ is the single-particle density matrix
(set by the Fermi distribution in the equilibrium case we are looking
at). $\hat{Q}$ is the occupation operator of the single level, coupling
to the qubit, and $\nu$ is the interaction strength. The results
depend on the ratio $\nu/\gamma$, where $\gamma$ is the tunneling
rate for an electron to escape from the defect level. 

\begin{figure}
\includegraphics[width=1\columnwidth]{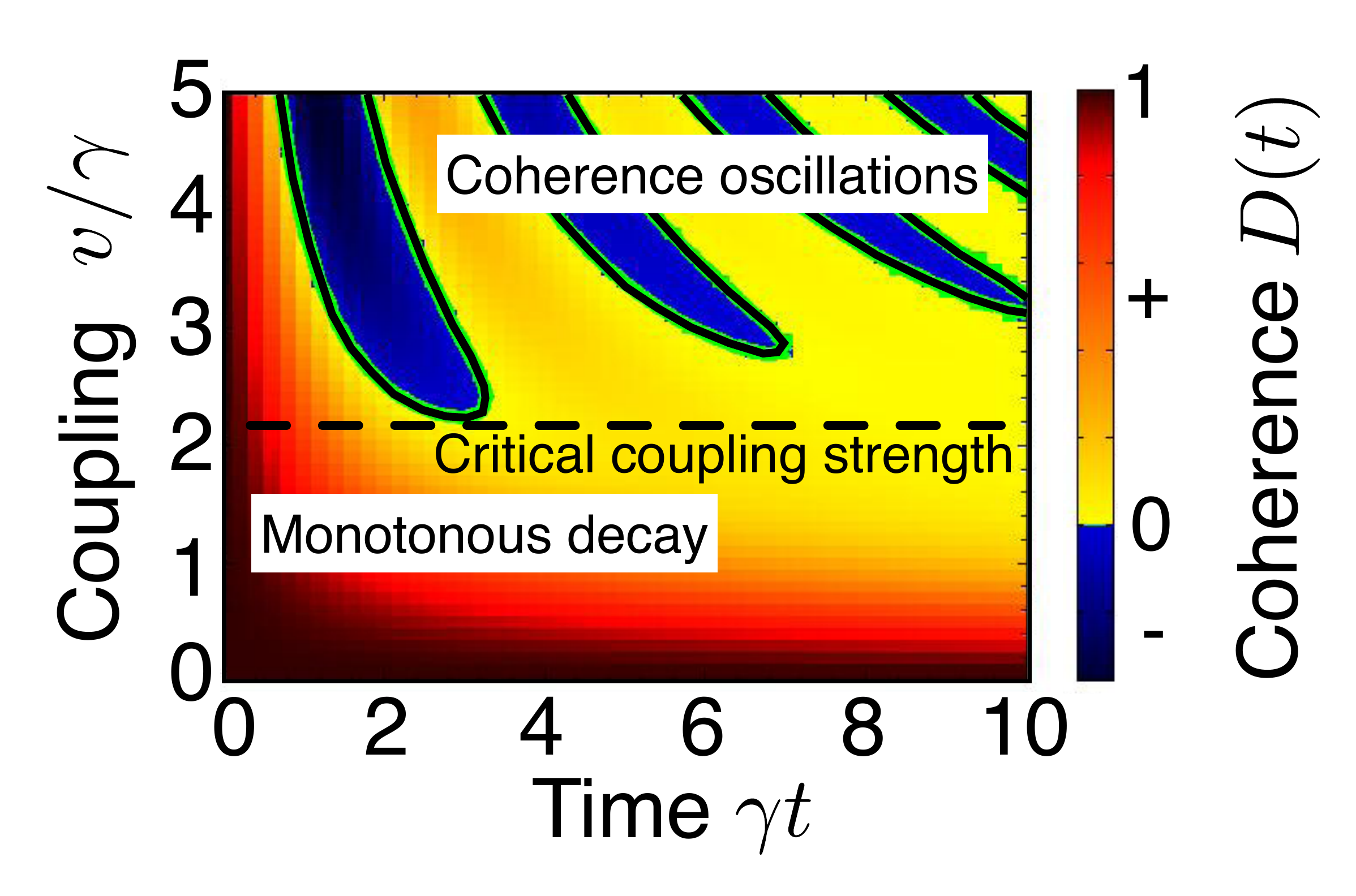}

\caption{\label{figquanttel}The coherence $D(t)$ for a qubit subject to quantum
telegraph noise (see Eq.~(\ref{eq:determinant}) and \cite{2008_05_AbelMarquardt_QuantumTelegraphNoise}),
for a low value of the temperature, $T/\gamma=0.2$, as a function
of time (horizontal) and coupling strength $\nu/\gamma$ (vertical).
$D(t)$ is displayed as a density plot, with the thick lines indicating
$D(t)\equiv0$. The fluctuator sits at $\epsilon=0$. (Note: $D(t)$
has been multiplied with a trivial phase factor to make it real-valued;
see \cite{2008_05_AbelMarquardt_QuantumTelegraphNoise} for more details
and related plots)}

\end{figure}

The classical result (\ref{PD-coherenceTelegraphNoise}) is recovered
in the high-temperature limit, justifying the name 'quantum telegraph
noise'. Even at lower temperatures, there is always a threshold in
$\nu/\gamma$ beyond which $D(t)$ displays the tell-tale oscillations
that characterize dephasing by non-Gaussian noise in the strong-coupling
regime. This is exemplified in Fig.~\ref{figquanttel}, where we
display $D(t)$ in the low-temperature limit. Thus, this model represents
one of the rare cases in which the non-trivial features of dephasing
by non-Gaussian fluctuations can be described exactly even deep in
the quantum regime.

\subsection{Decoherence by shot noise}

When current runs through a quantum wire, the resulting current or
charge fluctuations are due to discrete electrons and therefore non-Gaussian
(and non-equilibrium) in nature. We can construct a simple model,
relevant for charge-qubit readout, by postulating an interaction between
a qubit and the charge fluctuations in some section of a nearby 1d
ballistic quantum wire. Some current (driven by an applied voltage
$V$) is injected into the wire through a beam splitter with transmission
probability $\mathcal{T}$. 

Our analysis of this model \cite{2007_05_NederMarquardt_NJP_NonGaussian}
builds on the exact evaluation of a determinantal expression similar
to Eq.~(\ref{eq:determinant}). It reveals several interesting features:
(i) In such a 1d system, the charge fluctuations in equilibrium (i.e.
for zero voltage) are Gaussian, and only non-zero voltages will produce
interesting deviations from Gaussian dephasing. (ii) The expression
for the qubit's coherence is closely related to full-counting statistics,
and measuring its time-evolution may give access to that statistics
at intermediate times, different from the usual long-time limit. (iii)
For not too small interaction times, the results are essentially a
function of $eVt$, where $V$ is the applied voltage. (iv) Again
we find a certain ($V$-independent) threshold in the interaction
strength, beyond which coherence oscillations set in. 

A variant of this model has been realized recently in experiments:
By coupling the quantum wire capacitively to one channel of an electron
interferometer, it may be employed as a which-path detector. This
concept has been implemented in the Heiblum group at the Weizmann
Institute, making use of the electronic Mach-Zehnder interferometer
(see below). The qubit's coherence is then replaced by the interferometer's
interference contrast (visibility). Since the coupling between two
adjacent edge channels in the Quantum Hall effect can become very
large, that experiment has (apparently for the first time) entered
a regime where the visibility oscillations associated to dephasing
by non-Gaussian noise have been observed in an electronic interference
experiment \cite{2006_07_MZ_DephasingNonGaussianNoise_NederMarquardt}.

\section{Dephasing in fermionic systems}

\subsection{The electronic Mach-Zehnder interferometer}

Whereas the physics of decoherence has been studied in great detail
for single particle systems, much remains to be learned in dealing
with decoherence in many-particle systems. We refer the reader to \cite{2006_04_DecoherenceReview}
for a brief review.  In the past, we have studied several models to tackle the questions associated with
this topic, including: A ballistic ring containing many electrons, coupled to a fluctuating quantum flux\cite{2002_Marquardt_AB_PRB}, transport through a double dot interferometer subject to a noisy quantum bath \cite{2003_Marquardt_DD_Interferometer}, a many-fermion generalization of the Caldeira-Leggett model \cite{2004_Marquardt_DFS_PRL,2004_Marquardt_DFS_LongVersion}, decoherence by external quantum noise in an electronic Mach-Zehnder interferometer \cite{2004_10_Marquardt_MZQB_PRL}, and decoherence in weak localization \cite{2005_10_WeaklocDecoherenceOne}.

\begin{figure}
\includegraphics[width=1\columnwidth]{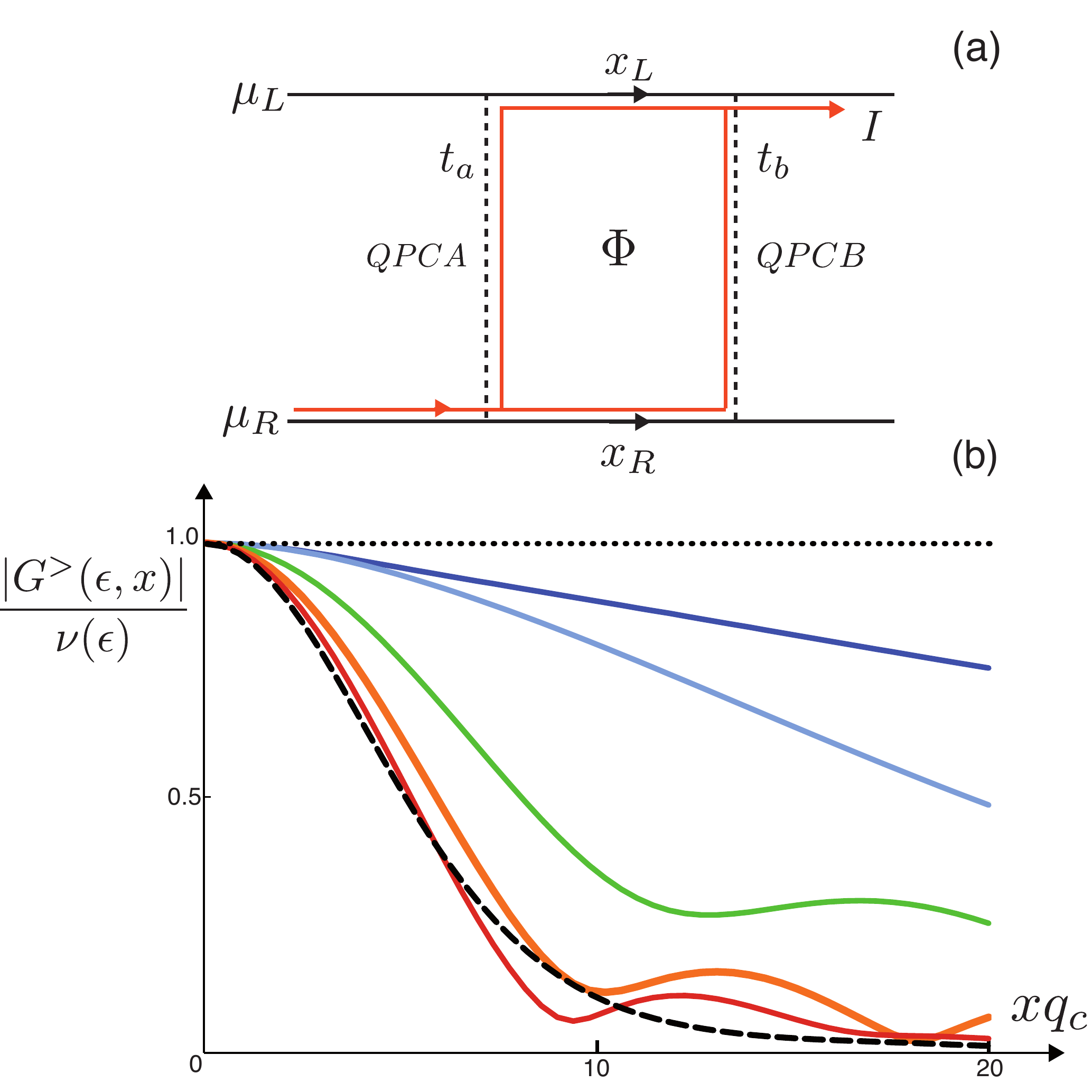}
\caption{\label{luttingerfig}(a) Scheme of the Mach-Zehnder interferometer.
The two channels $1$ and $2$ of length $x_{L,R}$ and the corresponding
chemical potentials $\mu_{L,R}$ are indicated. The electrons can
tunnel at QPCs A and B, with tunnel amplitudes $t_{a}$ and $t_{b}$.
By tuning the magnetic flux $\Phi$ through the interferometer, one
observes an interference pattern $I(\phi)$. (b) The electron's coherence,
$G^{>}(\epsilon,x)/\nu(\epsilon)$, as a function of propagation distance,
for various values of the energy $\epsilon$ (from top to bottom):
$\epsilon/(q_{c}v_{F})=0,0.54,1.15,1.75,2.36,2.96$. The dashed, black line
indicates the limit $\epsilon\rightarrow\infty$. Here, the interaction
potential is $U_{q}=2\pi\alpha v_{F}\exp(-(q/q_{c})^{2})$, with $\alpha=3.0/2\pi$.
\label{Fig1}}

\end{figure}

Of particular interest are electronic systems such as they occur in solid state
transport interference experiments. Especially the first realization
of the electronic Mach-Zehnder interferometer (MZI) (Fig.\ref{Fig1})
in the Heiblum group and subsequent experiments \cite{2003_Heiblum_MachZehnder,2006_01_Neder_VisibilityOscillations,2008_02_Strunk_MZ_Coherence_FillingFactor,2008_03_Roche_CoherenceLengthMZ}
provide a particularly beautiful clean model system in which to study
the effects of interactions and decoherence on the interference contrast in a many-fermion system .

Experimentally, the MZI is realized by making use of integer quantum
Hall edge channels, which are connected via quantum point contacts
(QPCs), representing the beam splitters in the well known optical
version of the interferometer. For our purposes, we will model this
by assuming that there are two chiral electronic channels, being weakly
tunnel-coupled to each other at two locations\textbf{ }with tunneling
amplitudes $t_{a}$ and $t_{b}$, respectively. Thus, we restrict
our discussion to highly reflecting QPCs in order to employ perturbation
theory (following \cite{2006_09_Sukhorukov_MZ_CoupledEdges,2007_ChalkerGefen_MZ}).
Applying a finite bias voltage $V$ between the channels leads to
a current $I$ through the interferometer. There is an Aharonov-Bohm
phase difference, such that $I(\Phi)$ becomes an oscillating function
of the magnetic flux $\Phi$. Introducing the maximum and the minimum
current with respect to $\Phi$, one may characterize the interference
contrast via the so-called visibility $\mathcal{V}\equiv(I_{{\rm {\rm max}}}-I_{{\rm min}})/(I_{{\rm {\rm max}}}+I_{{\rm min}})$.
The visibility can be used as a direct measure for the coherence of
the system. 

To take into account interactions in the chiral channels, one can
employ the technique of bosonization. Then one only needs to obtain
the electron Green's function (GF) in channel $j\in(L,R)$, $G_{j}^{>}(x,t)\equiv-i\left\langle \hat{\Psi}(x,t)\hat{\Psi}^{\dagger}(0,0)\right\rangle $
and the corresponding hole GF $G^{<}(x,t)$. We can find 
\cite{2008_06_Neuenhahn_UniversalDephasing_Short}
a rather intuitive expression for the visibility in terms of the GFs'
Fourier transform with respect to $t$ (here $T=0$):\begin{equation}
\mathcal{V}=\frac{2|t_{a}t_{b}^{\ast}|}{|t_{a}|^{2}+|t_{b}|^{2}}\cdot\frac{\left|\int_{0}^{\delta\mu}d\epsilon\; G_{L}^{>}(x_{L},\epsilon)G_{R}^{<}(-x_{R},\epsilon-\delta\mu)\right|}{\int_{0}^{\delta\mu}d\epsilon\;\nu_{L}(\epsilon)\nu_{R}(\epsilon-\delta\mu)}\,,\end{equation}
where $G^{<}(x,\epsilon)=G^{>*}(x,-\epsilon)$ and \\ $\nu_{L,R}(\epsilon)=\left|G_{L,R}^{>(<)}(0,\epsilon)\right|$
for $\epsilon>0$ ($\epsilon<0$). There are contributions from all
electrons inside the voltage interval, $\epsilon=0\ldots\delta\mu$,
where $\delta\mu=q_{e}V>0$ is the bias between the left and the right
arm of the interferometer. The propagation distance between the QPC's
in channel $j$ is denoted by $x_{j}$. 

Obviously, the visibility only depends on a product of the electron-
and the hole-GF. In order to understand the reason for this generic
feature, one has to think about the nature of coherence in many body
systems \cite{2006_04_DecoherenceReview,2005_10_WeaklocDecoherenceOne}.
Let us think of an electron starting, for example in channel $R$
(Fig.~\ref{luttingerfig}). In the end we measure the current at
the output port in channel $L$. When the electron arrives at QPC
A it can be reflected, remaining in channel $R$, propagating there
and finally tunneling into channel $L$, where it gets measured. In
addition, the electron can immediatly tunnel at the first QPC, leaving
a hole behind. After the propagation through channel $L$ the electron
is counted at the output port. Loosely speaking, the action of QPC
A turns the full electronic many-body state into a coherent superposition
of many-body states, which are characterized by the status of all
the electrons and holes in the system. Therefore scattering the hole
destroys the coherent superposition just as well as scattering the
electron itself. 

Note that the coherence of the electron (hole) is completly encoded
in the GF $G^{>(<)}(\epsilon,x)$. It yields the amplitude of an electron
at energy $\epsilon$ propagating from $x=0$ to $x>0$.

\subsection{\textit{\emph{Decoherence in chiral electron liquids}}}

When applying the general discussion to a chiral 1d electron system,
we first note that interactions lead to non-trivial effects only if
one allows the interaction potential $U(x)$ to have some finite range,
as emphasized in \cite{2007_ChalkerGefen_MZ}. Given the potential's
Fourier components, $U_{q}\equiv\int dx\, e^{-iqx}U(x)$, we introduce
the dimensionless coupling strength $\alpha=\frac{U(q\rightarrow0)}{2\pi v_{F}}$
(here $\hbar\equiv1$). Besides that, we just assume the potential
to fall off on a scale $q_{c}$ in momentum space.

In Fig.~\ref{luttingerfig}, we show an example of $G^{>}(\epsilon,x)$
as a function of $x$ for different energies $\epsilon$, obtained
by numerically evaluating the exact bosonization solution. One can
observe that it decays for increasing propagation distance $x$, reflecting
the decoherence of the electron moving through the channel. Obviously
the strength of the decoherence depends on the energy $\epsilon$.
Two different energy regimes show up: For energies close to the Fermi
edge, $\epsilon-\epsilon_{F}\ll q_{c}v_{F}$, the decoherence is suppressed,
becoming stronger at larger energies. Thus the visibility $\mathcal{V}$
decays upon increasing the bias voltage $V$. In the limit of high
energies, the coherence $\left|G^{>}(\epsilon,x)\right|$ becomes
energy-independent. 

In a recent work \cite{2008_06_Neuenhahn_UniversalDephasing_Short},
we have shown that one can apply a semiclassical approximation which gets
exact in the limit of high-energy electrons and provides a very intuitive
picture. The basic idea postulated in that work is a variant of the
equation-of-motion approach to decoherence in fermionic systems \cite{2004_10_Marquardt_MZQB_PRL,2006_04_MZQB_Long,2006_04_DecoherenceReview},
and is related to functional bosonization: The propagating electron
picks up a random phase originating from the potential fluctuations
due to all the other electrons. 

An unexpected nontrivial result of the analysis in \cite{2008_06_Neuenhahn_UniversalDephasing_Short}
is that the high-energy electrons display a universal power-law decay
of the coherence, i.e. $|G^{>}(\epsilon,x)|\propto1/x$ for $x\rightarrow\infty$,
with an exponent $1$ \emph{independent of interaction strength} (at
$T=0$). This can be connected to the universal low-frequency behaviour
of the fluctuation spectrum that is observed in the electron's moving
frame of reference. (To avoid confusion, note that $|G^{>}(\epsilon,x)|\equiv1$
in the absence of interactions!) Such a behaviour is in contrast to
the \emph{low-energy} (!) power-law decay of coherence of a non-chiral
Luttinger liquid, where the exponent does depend on the interaction
strength. Note that in the situation discussed here there is no decay
at low energies, since we are dealing with a chiral Fermi liquid.

\section{Conclusions}

In this brief review, we have recounted some of our recent developments
in the field of decoherence. Obviously, there are many problems that
remain to be analyzed in the future. These include treating non-Gaussian
noise in situations where only fully numerical methods may be applicable,
and dealing with interacting electronic interferometers without assumptions
like small tunnel coupling.

\emph{Acknowledgements}. - We would like to acknowledge support through
SFB/TR12, NIM, the Emmy-Noether program, and SFB 631. 

\bibliographystyle{pss}
\bibliography{/Users/florian/pre/bib/BibFM}

\end{document}